\documentclass[12pt]{article}
\usepackage{enumerate}
\usepackage{amsfonts}
\usepackage{amsmath}
\usepackage{amssymb}
\usepackage{amsthm}

\def\A{\mathcal{A}}
\def\B{\mathcal{B}}
\def\D{\mathcal{D}}
\def\H{\mathcal{H}}
\def\K{\mathcal{K}}
\def\F{\mathcal{F}}
\def\E{\mathcal{E}}
\def\L{\mathcal{L}}
\def\S{\mathfrak{S}}
\def\T{\mathfrak{T}}
\def\B{\mathfrak{B}}
\def\J{\mathfrak{J}}
\def\M{\mathfrak{M}}
\def\N{\mathfrak{N}}

\newcounter{defin}  \newcounter{lemma}  \newcounter{theorem}
\newcounter{property} \newcounter{corol}  \newcounter{remark} \newcounter{example}

\newenvironment{lemma}{\par\refstepcounter{lemma}
     \textbf{Lemma \thelemma.} }{\rm\par}
\newenvironment{theorem}{\par\refstepcounter{theorem}
     \textbf{Theorem \thetheorem.}\ }{\rm\par}
\newenvironment{property}{\par\refstepcounter{property}
     \textbf{Proposition \theproperty.}\ }{\rm\par}
\newenvironment{corollary}{\par\refstepcounter{corol}
     \textbf{Corollary \thecorol.} }{\rm\par}
\newenvironment{definition}{\par\refstepcounter{defin}
     \textbf{Definition \thedefin.}\ }{\rm\par}
\newenvironment{remark}{\par\refstepcounter{remark}
     \textbf{Remark \theremark.}}{\rm\par}
\newenvironment{example}{\par\refstepcounter{example}
     \textbf{Example \theexample.}}{\rm\par}

\newcommand{\id}{\mathrm{Id}}
\newcommand{\Tr}{\mathrm{Tr}}

\begin{document}
\title{Entropy reduction of quantum measurements\thanks{The work was done during a visit to the Institut Mittag-Leffler (Djursholm,
Sweden).}}

\author{M.E.~Shirokov\thanks{e-mail:msh@mi.ras.ru}\\\\
Steklov Mathematical Institute, Moscow, Russia}

\date{} \maketitle

\begin{abstract}
It is observed that the entropy reduction (the information gain in the initial terminology) of an efficient (ideal or pure) quantum measurement coincides with the generalized quantum mutual information of a q-c channel mapping an a priori state to the corresponding posteriori probability distribution of the outcomes of the measurement.  This observation makes it possible to define the entropy reduction for  arbitrary a priori states (not only for states with finite von Neumann entropy) and to study its analytical properties by using general properties of the quantum mutual information.

By using this approach one can show that the entropy reduction of an efficient quantum measurement is a nonnegative lower semicontinuous concave function on the set of all a priori states having continuous restrictions to subsets on which the von Neumann entropy is continuous. Monotonicity and subadditivity of the entropy reduction are also easily proved by this method.

A simple continuity condition for the entropy reduction and for the mean posteriori entropy considered as functions of a pair (a priori state, measurement) is obtained.

A characterization of an irreducible measurement (in the Ozawa sense) which is not efficient is considered in the Appendix.
\end{abstract}


\tableofcontents

\section{Introduction}

The notion of a quantum measurement plays a key role in quantum theory. One of quantitative characteristics of a quantum measurement is the entropy reduction\footnote{It was originally called the information gain (cf.\cite{G-IG,L-IG,O-IG}) but then the terminology had been changed (some arguments explaining this change can be found in \cite{BHH}).} defined as a difference between the von Neumann entropy of an a priori (pre-measurement) state and the mean von Neumann entropy of the corresponding posteriori (post-measurement) states. Roughly speaking, the entropy reduction characterizes a degree of purifying ("gain in purity") of a state in a measurement process. More details about the information sense of this value can be found in \cite{J,Luo,O-IG}.

An interesting question concerns the sign of the entropy reduction. Groenewold
has conjectured in \cite{G-IG} and Lindblad has proved in \cite{L-IG} that the entropy reduction is
nonnegative for quantum measurements of the von Neumann-Luders type. The general case has
been studied by Ozawa, who has proved in \cite{O-IG} that the entropy reduction is
nonnegative if and only if the quantum measurement is \emph{quasicomplete} (also called \emph{irreducible} in \cite{O-SR}) in the sense that for an arbitrary pure a priori state the corresponding family of posteriori states consists of pure states (for almost all outcomes). The class of  quasicomplete (irreducible) quantum measurements contains the class of \emph{efficient} or \emph{pure} measurement (cf. \cite{H-RND,J,Luo}) described in Section 3.1. Quantum measurements belonging to the gap between these two classes are characterized in the Appendix as measurements with quite singular properties.

In this paper we show that the entropy reduction of an efficient (pure) quantum measurement can be expressed via the (generalized) quantum mutual information of the quantum-classical channel mapping an a priori state to the corresponding posteriori probability distribution of the outcomes of the measurement. This makes it possible to define the entropy reduction for arbitrary a priori states (not only for states with finite von Neumann entropy) and to study its analytical properties by using results concerning
the quantum mutual information of infinite-dimensional channels \cite{H-Sh-3}.

The paper is organized as follows. In the first part we restrict attention to the case of quantum measurements with a discrete set of outcomes, which is more simple mathematically. In the second part we consider the case of general quantum measurements described by completely positive instruments by generalizing the notion of the quantum mutual information for channels taking values in the space of normal states of an arbitrary $W^*$-algebra.

\section{The discrete case}

Let $\mathcal{H}$ be a separable Hilbert space,
$\mathfrak{B}(\mathcal{H})$ -- the Banach space of all bounded
operators in $\mathcal{H}$ with the operator norm $\Vert\cdot\Vert$,
$\mathfrak{T}( \mathcal{H})$ -- the Banach space of all
trace\nobreakdash-\hspace{0pt}class operators in $\mathcal{H}$ with
the trace norm $\Vert\cdot\Vert_{1}=\Tr|\cdot|$, containing the cone
$\mathfrak{T}_{+}(\mathcal{H})$ of all positive
trace\nobreakdash-\hspace{0pt}class operators. The closed convex
subset
$\mathfrak{S}(\mathcal{H})=\{A\in\mathfrak{T}_{+}(\mathcal{H})\,|\,\mathrm{Tr}A=1\}$
is a complete separable metric space with the metric defined by the
trace norm. Operators in $\mathfrak{S}(\mathcal{H})$ are denoted
$\rho,\sigma,\omega,...$ and called density operators or states
since each density operator uniquely defines a normal state on
$\mathfrak{B}(\mathcal{H})$ \cite{B&R}. \smallskip

The identity operator in a Hilbert space $\mathcal{H}$
and the identity transformation of the set $\T(\H)$ will be denoted $I_{\H}$ and $\id_{\H}$
correspondingly. \smallskip

For an arbitrary state $\omega\in\S(\H\otimes\K)$ the partial states
$\Tr_{\K}\omega$ and $\Tr_{\H}\omega$
will be denoted
$\,\omega_{\H}$ and $\,\omega_{\K}$.\smallskip

We will use the
following natural extension of the von Neumann entropy $H(\rho)=-\Tr\rho\log\rho$ of a quantum state
$\rho\in\mathfrak{S}(\mathcal{H})$ to the cone
$\mathfrak{T}_{+}(\mathcal{H})$ of all positive
trace\nobreakdash-\hspace{0pt}class operators
$$
H(A) =\Tr AH\left(\frac{A}{\Tr A}\right)=\Tr \eta (A) - \eta(\Tr A),\quad\eta(x)=-x\log
x.\footnote{$\log$ denotes the natural logarithm.}
$$

The quantum relative entropy is defined for
arbitrary operators $A$ and $B$ in $\mathfrak{T}_{+}(\mathcal{H})$
as follows
\begin{equation}\label{qre}
H(A\,\|B)=\sum_{i}\langle i|\,(A\log A-A\log B+B-A)\,|i\rangle,
\end{equation}
where $\{|i\rangle\}$ is the orthonormal basis of eigenvectors of
$A$ and it is assumed that $H(A\,\|B)=+\infty$ if $\,\mathrm{supp}A$
is not contained in $\mathrm{supp}B$ \cite{L-2}. \smallskip

A linear completely positive trace\nobreakdash-\hspace{0pt}preserving map
$\Phi:\mathfrak{T}(\mathcal{H})\rightarrow\mathfrak{T}(\mathcal{H}')$
is called a \textit{quantum channel}
\cite{H-SSQT,N&Ch}.

By the
Stinespring dilation theorem there exist a separable Hilbert space
$\mathcal{H}''$ and an isometry
$V:\mathcal{H}\rightarrow\mathcal{H}'\otimes\mathcal{H}''$ such that
\begin{equation}\label{Stinespring-rep}
\Phi(A)=\mathrm{Tr}_{\mathcal{H}'}VA V^{*},\quad \forall
A\in\mathfrak{T}(\mathcal{H}).
\end{equation}
The quantum channel
\begin{equation}\label{c-channel}
\mathfrak{T}(\mathcal{H})\ni
A\mapsto\widetilde{\Phi}(A)=\mathrm{Tr}_{\mathcal{H}''}VAV^{*}\in\mathfrak{T}(\mathcal{H}'')
\end{equation}
is called \emph{complementary}\footnote{The channel
$\widetilde{\Phi}$ is also called \emph{conjugate} or
\emph{canonically dual} to the channel $\Phi$ \cite{KMNR,Winter}.} to the channel $\Phi$, it is uniquely
defined up to unitary equivalence \cite{H-c-c}. \smallskip

The \emph{quantum mutual information} is an important entropic characteristic of a channel $\Phi:\T(\H)\rightarrow\T(\H')$ related to the entanglement-assisted classical capacity of this channel \cite{H-SSQT, N&Ch}. In finite dimensions it is defined at arbitrary state $\rho\in\mathfrak{S}(\H)$ by the expression (cf.\cite{A&C})
\begin{equation}\label{mi}
I(\rho, \Phi)=H(\rho)+H(\Phi(\rho))-H(\widetilde{\Phi}(\rho)).
\end{equation}
In infinite dimensions this expression may contain uncertainty $"\infty-\infty"$, but it
can be modified to avoid this problem as follows
\begin{equation}\label{mi+}
I(\rho, \Phi) = H\left(\Phi \otimes \id_{\K}
(\hat{\rho}) \| \Phi \otimes \id_{\K} (\rho\otimes
\varrho\,)\right),
\end{equation}
where $\K$ is a Hilbert space isomorphic to $\H$,
$\,\hat{\rho}\,$ is a purification\footnote{This means that
$\Tr_{\K}\hat{\rho}=\rho$.} of the state $\rho$ in the space $\H\otimes\K$ and $\varrho=\Tr_{\H}\hat{\rho}$ is a state in $\S(\K)$ isomorphic to $\rho$. Analytical properties of the function $(\rho, \Phi)\mapsto I(\rho, \Phi)$ defined by (\ref{mi+}) in the infinite dimensional case are studied in \cite{H-Sh-3}. \smallskip

An efficient quantum measurement $\M$ with a countable outcome set $X=\{x_i\}_{i\in I}$ is described by a set $\{V_i\}_{i\in I}$ of operators in $\B(\H)$ such that
$\sum_{i\in I}V_i^*V_i=I_{\H}$. Applying this measurement to an arbitrary a priori state
$\rho\in\mathfrak{S}(\H)$ results in the posteriori probability distribution $\{\pi_i(\rho)\}_{i\in I}$, where
$\pi_i(\rho)=\Tr V_i\rho V^*_i$ is the probability of the outcome $x_i$, and the corresponding family
of posteriori states $\{\rho_i\}_{i\in I}$, where $\rho_i=(\pi_i(\rho))^{-1}V_i\rho V^*_i$.
Thus $\sum_{i\in I}\pi_i(\rho)H(\rho_i)=\sum_{i\in I}H(V_i\rho V^*_i)$ is the mean entropy of posteriori states.
The \emph{entropy reduction} of the quantum measurement $\M$ at an a priori state $\rho$ with finite entropy is the following value
$$
ER(\rho,\M)\doteq H(\rho)-\sum_{i\in I}\pi_i(\rho)H(\rho_i)=H(\rho)-\sum_{i\in I}H(V_i\rho V^*_i).
$$

Let $\H_X$ be a Hilbert space having dimension coinciding with the cardinality of the outcome set $X$. Consider the quantum channel
\begin{equation}\label{q-c-ch}
\mathfrak{T}(\mathcal{H})\ni A \mapsto\Pi_{\M}(A)=\sum_{i\in I}\mathrm{Tr}\left[V_{i}AV_{i}^{*}\right]|\varphi_i\rangle\langle
\varphi_i|\in \mathfrak{T}(\mathcal{H}_X),
\end{equation}
where $\{|\varphi_i\rangle\}_{i\in I}$ is a particular
orthonormal basis in $\H_X$. The channel $\Pi_{\M}$ is a quantum modification of the quantum-classical channel mapping a state $\rho$ to the probability distribution $\{\pi_i(\rho)\}_{i\in I}$. It is essential that
\begin{equation}\label{b-eq}
ER(\rho,\M)=I(\rho,\Pi_{\M})
\end{equation}
for any state $\rho$ in $\mathfrak{S}(\H)$ with finite entropy.
If $H\left(\{\pi_i(\rho)\}_{i\in I}\right)<+\infty$ this equality
directly follows from (\ref{mi}), since $\widetilde{\Pi}_{\M}(\cdot)=\sum_{i\in I}U_iV_i(\cdot)V^*_iU^*_i$, where $\{U_i\}_{i\in I}$ is a family of isometrical embedding of $\H$ into $\bigoplus_{i\in I}\H_i$, $\H_i\cong\H$, such that $U_i\H=\H_i$. In general case it
can be easily deduced\footnote{It can be proved directly by the obvious modification of the proof of Proposition \ref{b-eq+} in Section 3.2.} from \cite[Proposition 3 and Theorem 1]{H-Sh-3}.\smallskip

Equality (\ref{b-eq}) obtained under the condition $H(\rho)<+\infty$ makes it possible to consider the entropy reduction of an efficient quantum measurement with a countable outcome set as a function on the whole space of a priori states.  \medskip

\begin{definition}\label{QIG}
The entropy reduction of an efficient quantum measurement $\M=\{V_i\}_{i\in I}$  at an \emph{arbitrary} a priori state $\,\rho\,$ is
defined as follows
$$
ER(\rho,\M)\doteq I(\rho, \Pi_{\M}),
$$
where $\,\Pi_{\M}\,$ is the quantum channel defined by (\ref{q-c-ch}).
\end{definition}\medskip

By equality (\ref{b-eq}) this definition is consistent with the conventional one.
Its main advantage consists in possibility to study the function $\rho\mapsto ER(\rho, \M)$ on the whole space of a priori states
by using properties of the quantum mutual information (many of them follow from the corresponding properties of the quantum relative entropy).\smallskip

\begin{theorem}\label{QIG-p}
\emph{Let $\,\M$ be an efficient measurement in a  Hilbert space $\H$ with a countable outcome set. The function $\,\rho\mapsto ER(\rho,\M)$ is nonnegative concave and lower semicontinuous on the set $\,\S(\H)$. It has the following properties:}
\begin{enumerate}[1)]
    \item \emph{$\{ER(\rho,\M)=0\}\Leftrightarrow\{\rho_i\cong\rho\;\,\forall i\;\,s.t.\;\pi_i(\rho)\neq0\}$,
where $\{\pi_i(\rho)\}$ and $\{\rho_i\}$ are respectively the posteriori probability distribution and the family of posteriori states corresponding to an a priori state $\rho$;}

    \item \emph{continuity on any subset of $\,\S(\H)$ on which the von Neumann entropy is continuous:
    $$
\lim_{n\rightarrow+\infty}
H(\rho_n)=H(\rho_0)<+\infty\;\,\Rightarrow\;\,
\lim_{n\rightarrow+\infty}ER(\rho_n,\M)=ER(\rho_0,\M)<+\infty
$$
for any sequence $\{\rho_n\}$ of states converging to a state $\rho_0$;}
\item \emph{monotonicity: for arbitrary efficient measurements $\,\M=\{V_i\}_{i\in I}$ and $\,\N=\{U_j\}_{j\in J}$ in a  Hilbert space $\H$ with the outcome sets $\,X$ and $\,Y$ the inequality
\begin{equation}\label{monoton}
ER(\rho, \N\circ\M)\geq ER(\rho, \M)
\end{equation}
holds for any $\,\rho\in\mathfrak{S}(\H)$, where $\,\N\circ\M$ is the measurement in the space $\H$ with the outcome set $\,X\times Y$ determined by the family $\,\{U_jV_i\}_{i\in I, j\in J}$;}

\item \emph{subadditivity: for arbitrary efficient measurements $\;\M=\{V_i\}_{i\in I}$ and $\;\N=\{U_j\}_{j\in J}$ in Hilbert spaces $\H$ and $\K$ with the outcome sets $\,X$ and $\,Y$ the inequality
\begin{equation}\label{subaddit}
ER(\omega,\M\otimes\N)\leq ER(\omega_{\H},\M)+ER(\omega_{\K},\N)
\end{equation}
holds for any $\omega\in\mathfrak{S}(\H\otimes\K)$, where $\,\M\otimes\N$ is the measurement in the space $\H\otimes\K$ with the outcome set $\,X\times Y$ determined by the family $\,\{V_i\otimes U_j\}_{i\in I, j\in J}$.}

\end{enumerate}

\end{theorem}\vspace{5pt}

\textbf{Proof.} 1) Note that the equality
$$
ER(\rho,\M)\doteq I(\rho,\Pi_{\M})\doteq H(\Pi_{\M}\otimes \id_{\K}
(\hat{\rho}) \| \Pi_{\M}(\rho) \otimes
\varrho)=0,
$$
where $\hat{\rho}$ is a purification of $\rho$ and $\varrho=\Tr_{\H}\hat{\rho}$, means
\begin{equation}\label{s-eq}
\Pi_{\M}\otimes \id_{\K}
(\hat{\rho}) = \Pi_{\M}(\rho) \otimes
\varrho
\end{equation}
by the well known property of the relative entropy.

Let $\rho =
\sum_{k}\lambda_k |k\rangle\langle
k|$ and $\hat{\rho} = \sum_{j,k}\sqrt{\lambda_j}\sqrt{\lambda_k}
|j\rangle\langle
k| \otimes |j\rangle\langle
k|$. It is easy to see that (\ref{s-eq}) is equivalent to
\begin{equation}\label{is}
\langle
k|V_i^*V_i|j\rangle=\delta_{jk}\Tr V_i^*V_i\rho=\delta_{jk}\pi_i(\rho)\quad \textup{for}\; \textup{all}\;\; i,j,k.
\end{equation}

$"\Leftarrow"$ \footnote{$\;"\Leftarrow"$ is obvious only if the state $\rho$ has finite entropy.}
Let $\varrho_i=(\pi_i(\rho))^{-1}\Tr_{\H}V_i\otimes I_{\K}\cdot\hat{\rho}\cdot V^{*}_i\otimes I_{\K}$. Then $\varrho_i\cong\rho_i$ and hence $\varrho_i\cong\varrho$, since $\varrho\cong\rho$. By noting that $\varrho=\sum_{i\in I}\pi_i(\rho)\varrho_i$ we conclude from Lemma \ref{simple} in the Appendix that $\varrho_i=\varrho\,$ for all $i$. Since
$
\varrho_i=(\pi_i(\rho))^{-1}\sum_{j,k}\sqrt{\lambda_j}\sqrt{\lambda_k}[\Tr V_i|j\rangle\langle
k|V_i^*]|j\rangle\langle k|$ and
$\varrho =
\sum_{k}\lambda_k |k\rangle\langle
k|$,
we obtain (\ref{is}).\smallskip

$"\Rightarrow"$ Relations (\ref{is}) mean that  $PV_i^*V_iP=\pi_i(\rho)P$ for each $i$, where $P=\sum_{k}|k\rangle\langle
k|$ is the projector on the support of the state $\rho$. Thus $(\pi_i(\rho))^{-1/2}V_iP$ is a partial isometry and hence
$\rho_i=(\pi_i(\rho))^{-1}V_iP\rho PV_i^*\cong\rho$
for each $i$ such that $\pi_i(\rho)\neq0$.\smallskip

2) This directly follows from Proposition 4 in \cite{H-Sh-3}.

3) This follows from the 1-st chain rule for the quantum mutual information (property 3 in Proposition 1 in \cite{H-Sh-3}). Indeed,
$$
\Pi_{\N\circ\M}(A)=\sum_{i\in I,j\in J}\Tr\left[V_{i}^{*}U_{j}^{*}U_{j}V_{i}A\right]|i\rangle\langle
i|\otimes|j\rangle\langle
j|,\quad A\in\,\mathfrak{T}(\mathcal{H}),
$$
where $\{|i\rangle\}_{i\in I}$ and $\{|j\rangle\}_{j\in J}$ are particular
orthonormal bases in the spaces $\H_X$ and $\H_Y$, and hence $\Pi_{\M}(A)=\Tr_{\H_Y}\Pi_{\N\circ\M}(A)$.

4) This follows from subadditivity of the quantum mutual information (property 5 in Proposition 1 in \cite{H-Sh-3}), since
$\Pi_{\M\otimes\N}=\Pi_{\M}\otimes\Pi_{\N}$. $\square$ \medskip

Consider the question of continuity of the entropy reduction with respect to "perturbation" of quantum measurements. \smallskip

Let $\mathbb{M}(\H)$ be the set of all efficient quantum measurements in the Hilbert space $\H$ with finite or countable set of outcomes identified with the set of all sequences
$\{V_{i}\}_{i=1}^{+\infty}$ of operators in $\B(\H)$ such that
$\sum_{i=1}^{+\infty}V^{*}_{i}V_{i}=I_{\H}$ endowed with the topology
of coordinate\nobreakdash-\hspace{0pt}wise strong operator
convergence. Proposition 1 in \cite{H-Sh-3} and  Corollary 2 in \cite{H-Sh-3} imply the following assertion. \smallskip

\begin{property}\label{cont-cond}
\emph{The function
\begin{equation}\label{fun}
(\rho, \M)\mapsto ER(\rho, \M)
\end{equation}
is lower semicontinuous on the set
$\,\S(\H)\times\mathbb{M}(\H)$.}
\emph{Let $\,\mathcal{A}$ be an arbitrary subset
of $\,\mathfrak{S}(\mathcal{H})$ on which the von
Neumann entropy is continuous. Then function
(\ref{fun}) is continuous on the set
$\,\mathcal{A}\times\mathbb{M}(\H)$.}
\end{property}
\medskip

By Proposition \ref{cont-cond} and Proposition 6.6 in
\cite{O&P} function
(\ref{fun}) is continuous on the set
$\,\mathcal{K}_{H,h}\times\mathbb{M}(\H)$, where
$\mathcal{K}_{H,h}$ is the set of states with the mean energy $\Tr H\rho$ not exceeding $h>0$ provided the Hamiltonian $H$ of the
quantum system satisfies the condition $\Tr e^{-\lambda H}<+\infty$ for all $\lambda>0$ (which holds, for example, for the Hamiltonian of the system of quantum oscillators).

\section{The general case}

\subsection{On properties of efficient (pure) instruments}

A general quantum measurement in a Hilbert space $\H$ with the measurable outcome set $\{X,\F\}$
is described by a special mathematical object called \emph{instrument}, which was introduced by Davis and Lewis \cite{D&L}.
An instrument $\M$ (in a space of states) is a $\sigma$-additive measure on $\{X,\F\}$ taking values within the set of quantum operations -- completely positive trace-non-increasing linear transformations of $\T(\H)$ such that $\M(X)$ is a channel (see the detailed definition in \cite{B&L,H-SSQT,L}).

Let $\rho$ be an arbitrary a priori state in $\S(\H)$. Then the outcome of the measurement $\M$ is contained in a set $F\in\F$ with probability
$\Tr\M(F)[\rho]$. If this probability is nonzero then $\,(\Tr\M(F)[\rho])^{-1}\M(F)[\rho]\,$ is the corresponding posteriori state of the system.
Thus $\,F\mapsto\mu_{\rho}(F)\doteq\Tr\M(F)[\rho]\,$ is the posteriori probability measure on the outcome set $\{X,\F\}$ corresponding to the a priori state $\rho$.

Ozawa proved in \cite{O-PS} existence of a family $\{\rho_x\}_{x\in X}$ of posteriori states defined for $\mu_{\rho}$\nobreakdash-\hspace{0pt}almost all $x$ such that
the function $x\mapsto\Tr A\rho_x$ is $\F$-measurable for any $A\in\B(\H)$ and
$$
\int_F\rho_x\mu_{\rho}(dx)=\M(F)[\rho]\quad \forall F\in\F.\quad \textup{(Bochner integral)}
$$

By using the family $\{\rho_x\}_{x\in X}$ one can consider the mean entropy of posteriori state $\,\int_X H(\rho_x)\mu_{\rho}(dx)\,$
and assuming that $H(\rho)<+\infty$ one can define the entropy reduction as follows
$$
ER(\rho,\M)=H(\rho)-\int_X H(\rho_x)\mu_{\rho}(dx).
$$
This is a natural generalization of the entropy reduction considered in Section 2 for the class of measurements with a countable set of outcomes.

Ozawa proved in \cite{O-IG} that $ER(\rho,\M)$ is nonnegative if and only if the instrument $\M$ is irreducible in the sense of the following definition.\footnote{In \cite{O-IG} the term \emph{quasicomplete} is used. The term \emph{irreducible} seems to be more reasonable, it appeared in \cite{O-SR} and is used in the subsequent papers.}\smallskip
\begin{definition}\label{irr-instr-def}
An instrument $\M$ is called \emph{irreducible} if for an arbitrary pure a priori state $\rho$ the posteriori states $\rho_x$ are pure
for $\mu_{\rho}$-almost all $x$.
\end{definition}\medskip

An arbitrary instrument $\M$ in a Hilbert space $\H$ can be represented as follows\footnote{The map $\,\M^*(F):\B(\H)\rightarrow\B(\H)\,$ is a dual map to the map $\,\M(F):\T(\H)\rightarrow\T(\H)\,$.}
\begin{equation}\label{SR}
\M^*(F)[A]=V^{*}\cdot A\otimes P(F)\cdot V,\quad A\in\B(\H),
\end{equation}
where $V$ is an isometry from $\H$ into $\H\otimes\H_{0}$ and $P(F)$ is a spectral measure in $\H_{0}$ \cite{O-SR} (see also \cite{H-RND}). \smallskip

The following notion introduced in \cite{H-RND} is a natural generalization of the notion of an efficient measurement with a countable outcome set.\smallskip
\begin{definition}\label{pure-instr-def}
An instrument $\M$ is called \emph{efficient} or \emph{pure} if it has representation (\ref{SR}) with the spectral measure
$P(F)$ of multiplicity one.\footnote{Below we will use the term \emph{efficient} to be consistent with the accepted terminology.}
\end{definition}\medskip

In \cite{H-RND} it is shown that an efficient instrument is irreducible. The converse assertion is not true (see Example \ref{i-np-ch-e} in the Appendix). A characterization of a quantum instrument, which is irreducible but not efficient, and a simple sufficient condition providing equivalence of efficiency and irreducibility are  presented respectively in Proposition \ref{i-np-ch} and in Corollary \ref{i-np-ch-c} in the Appendix. This characterization shows that a quantum instrument, which is irreducible but not efficient, has quite singular properties.

 It is the class of efficient instruments to which the results of Section 2 can be extended. We will use the following characterization of such instruments, where $\M\otimes\J_{\K}$ denotes the instrument $\M(\cdot)\otimes\id_{\K}$ in the Hilbert space $\H\otimes\K$ with the outcome set of the instrument $\M$. \medskip

\begin{property}\label{pure-inst} \emph{Let $\,\M$ be an instrument in a Hilbert space $\H$ with the outcome set $\,\{X,\,\F\}$. The following statements are equivalent:}
\begin{enumerate}[(i)]
  \item \emph{the instrument $\,\M$ is efficient;}
  \item \emph{the instrument $\,\M\otimes\J_{\K}$ is irreducible for 2-D Hilbert space $\K$;}
  \item \emph{the instrument $\,\M\otimes\J_{\K}$ is irreducible for a separable Hilbert space $\K$;}
  \item \emph{there exists a positive $\sigma$-finite measure $\mu$ on $\,\{X,\,\F\}$, a dense domain $\D\subset\H$ and a function $\;x\mapsto V(x)\,$ defined for $\mu$-almost all $\,x$, such that $V(x)$ is a linear operator from $\D$ to $\H$, satisfying
      $$
      \langle\varphi|\,\M^*(F)[A]\varphi\rangle=\int_F\langle V(x)\varphi|AV(x)\varphi\rangle\mu(dx)
      $$
      for any $\,\varphi\in\D$, $\,F\in\F$, $\,A\in\B(\H)$.}
      \item \emph{statement $\mathrm{(iv)}$ holds with $\D=\mathrm{lin}\{|\varphi_i\rangle\}$,  where $\{|\varphi_i\rangle\}$ is a given arbitrary orthonormal basis in $\H$.}
\end{enumerate}
\end{property}

\textbf{Proof.} $\mathrm{(i)\Rightarrow(iii)}$. It is easy to see that the instruments $\M$ and $\,\M\otimes\J_{\K}$
have representation (\ref{SR}) with the same spectral measure. Hence efficiency of $\M$ is equivalent to efficiency of $\,\M\otimes\J_{\K}$. As mentioned after the proof of Theorem 1 in \cite{H-RND} any efficient instrument is irreducible.

$\mathrm{(iii)\Rightarrow(ii)}\,$ is obvious, since the instrument in $\mathrm{(ii)}$ can be considered as a restriction of the instrument in $\mathrm{(iii)}$.

$\mathrm{(ii)\Rightarrow(v)}$. By  Theorem 1 in \cite{H-RND} and its proof there exist a positive $\sigma$\nobreakdash-\hspace{0pt}finite measure $\mu$ on $\,\{X,\,\F\}$  and a countable family $\{x\mapsto V_k(x)\}_k$ of functions defined for $\mu$-almost all $x$, such that $V_k(x)$ is a linear operator from $\D=\mathrm{lin}(\{|\varphi_i\rangle\})$ to $\H$, satisfying
$$
\langle\varphi|\,\M^*(F)[A]\varphi\rangle=\int_F\sum_k\langle V_k(x)\varphi|AV_k(x)\varphi\rangle\mu(dx)
$$
for any $\varphi\in\D$, $F\in\F$, $A\in\B(\H)$. Consider the family $\{\widehat{V}_k(x)=V_k(x)\otimes\id_{\K}\}$ of linear operators from $\,\widehat{\D}=\mathrm{lin}(\{\varphi_i\otimes\phi_j\}_{ij})$ to $\H\otimes\K$, where $\{|\phi_j\rangle\}$ is an orthonormal basis of the space $\K$. By using the polarization identity it is easy to show that
$$
\langle\hat{\varphi}|\,\M^*(F)\otimes\id_{\K}[C]\,\hat{\varphi}\rangle=\int_F\sum_k\langle \widehat{V}_k(x)\hat{\varphi}|C\widehat{V}_k(x)\hat{\varphi}\rangle\mu(dx)
$$
for any $\hat{\varphi}\in\widehat{\D}$, $\,F\in\F$, $\,C\in\B(\H\otimes\K)$. Hence for the instrument $\,\M\otimes\J_{\K}$
and a pure a priori state $\hat{\rho}=|\hat{\varphi}\rangle\langle\hat{\varphi}|$ (where $\hat{\varphi}\in\widehat{\D}$)
we have
\begin{equation}\label{m-rel}
\mu_{\hat{\rho}}(dx)=\sum_k\|\widehat{V}_k(x)\hat{\varphi}\|^2\mu(dx)
\end{equation}
while the posteriori state corresponding to the outcome $\,x\in X\setminus X_{\hat{\varphi}}^s\,$ is
\begin{equation}\label{p-state}
\hat{\rho}_x=\frac{\sum_k|\widehat{V}_k(x)\hat{\varphi}\rangle\langle \widehat{V}_k(x)\hat{\varphi}|}{\sum_k\|\widehat{V}_k(x)\hat{\varphi}\|^2},
\end{equation}
where $X_{\hat{\varphi}}^s=\{x\in X\,|\,\sum_k\|\widehat{V}_k(x)\hat{\varphi}\|^2=0\}$ is a set such that $\mu_{\hat{\rho}}(X_{\hat{\varphi}}^s)=0\,$ \cite{H-RND}.

By noting that the instrument $\,\M\otimes\J_{\K}$ is irreducible and by using (\ref{m-rel}) one can show existence of a set $X_{\hat{\varphi}}\in\F$ such that $\mu(X\setminus X_{\hat{\varphi}})=0$ and the above state
$\hat{\rho}_x$ is pure for any $x\in X_{\hat{\varphi}}\setminus X_{\hat{\varphi}}^s$ .

Let $\widehat{\D}_0$ be a countable subset of $\widehat{\D}$ consisting of finite linear combinations of the vectors of the family $\{\varphi_i\otimes\phi_j\}_{ij}$ with rational coefficients. Let $X_0=\bigcap_{\hat{\varphi}\in\widehat{\D}_0} X_{\hat{\varphi}}\in\F$.
Then $\mu(X\setminus X_0)=0$ and the state
$\hat{\rho}_x$ defined by (\ref{p-state}) is pure for all $\hat{\varphi}\in\widehat{\D}_0$ and all $x\in X_0\setminus X_{\hat{\varphi}}^s$. Hence the family $\{|\widehat{V}_k(x)\hat{\varphi}\rangle\}_k$ consists of collinear vectors for all $\hat{\varphi}\in\widehat{\D}_0$ and all $x\in X_0$. Since the rank of the operator $\widehat{V}_k(x)$ is either $0$ or $>1$, Lemma \ref{la} in the Appendix shows that $\widehat{V}_k(x)=\lambda_k(x)\widehat{V}_1(x)$ and hence $V_k(x)=\lambda_k(x)V_1(x)$, where $\lambda_k(x)\in\mathbb{C}$, for all $k$ and all $x\in X_0$.

Consider the linear operator $\,V(x)=\displaystyle\sqrt{\textstyle\sum_k|\lambda_k(x)|^2}\, V_1(x)$ defined on the set $\D$ for all $x\in X_0$.
It is easy to see that
$$
\langle\varphi|\M^*(F)[A]\varphi\rangle=\int_F\langle V(x)\varphi|AV(x)\varphi\rangle\mu(dx)
$$
for any $\,\varphi\in\D$, $\,F\in\F$, $\,A\in\B(\H)$. \medskip

$\mathrm{(v)\Rightarrow(vi)}\,$ is obvious.\medskip

$\mathrm{(iv)\Rightarrow(i)}$. By the condition the linear operator
$$
\D\ni\varphi\mapsto V(x)\varphi\in L_2(X,\F,\mu,\H)\cong\H\otimes L_2(X,\F,\mu)
$$
is isometrical and hence it can be extended to the isometry $V$ from $\H$ into $\H\otimes L_2(X,\F,\mu)$. A direct verification shows that
$$
\langle\varphi|\M^*(F)[A]\varphi\rangle=\langle V\varphi|(A\otimes P(F))V\varphi\rangle=\langle\varphi|V^*(A\otimes P(F))V\varphi\rangle
$$
for any vector $\varphi$ in $\H$, where $P(\cdot)$ is the spectral measure defined as follows\break $(P(F)f)(x)=\chi_{F}(x)f(x)$ for any $f\in L_2(X,\F,\mu)$, where $\chi_{F}(\cdot)$ is the indicator function of the set $F\in\F$. Thus
the instrument $\M$ is efficient. $\square$ \medskip

We will also use the following simple observations.

\begin{lemma}\label{comp}
1) \emph{For arbitrary efficient instruments $\,\M$ and $\,\N$ in a separable Hilbert space $\H$ with the outcome sets $\,\{X, \F\}$ and $\,\{Y, \E\}$ the instrument $\,\N\circ\M$  in the space $\H$ with the outcome set $\,\{X\times Y,\, \F\otimes\E\}$, defined by the relation $\,\N\circ\M(F\times E)=\N(E)\circ\M(F)$, $F\in\F$, $E\in\E$, is efficient.}\smallskip

2) \emph{For arbitrary efficient instruments $\,\M$ and $\,\N$ in separable Hilbert spaces $\H$ and $\K$ with the outcome sets $\,\{X, \F\}$ and $\,\{Y, \E\}$ the instrument $\,\M\otimes\N$  in the space $\H\otimes\K$ with the outcome set $\,\{X\times Y,\, \F\otimes\E\}$, defined by the relation $\,\M\otimes\N(F\times E)=\M(F)\otimes\N(E)$, $F\in\F$, $E\in\E$, is efficient.}
\end{lemma}\medskip

\textbf{Proof.} It is easy to show that the instruments $\N\circ\M$ and $\M\otimes\N$ have representation (\ref{SR}) with the spectral measure $P_{\M}\otimes P_{\N}$, where $P_{\M}$ and $P_{\N}$ are spectral measures corresponding to the instruments $\M$ and $\N$. $\square$ \smallskip

\subsection{A representation of the entropy reduction}

To extend the results of the previous section to the case of general type measurement consider the construction proposed by Barchielli and Lupieri in \cite{B&L}. Choose a positive complete measure $\mu_{0}$ on $(X,\F)$ such that $\mu_{\rho}$ is absolutely continuous with respect to $\mu_{0}$ for all $\rho$ in $\S(\H)$ (this can be done by using the measure $\mu_{\rho_{0}}$, where $\rho_{0}$ is a given full rank state in $\S(\H)$). Let $L_\infty(X,\F,\mu_0,\B(\H))$ be the W*-algebra of $\mu_0$-essentially bounded $\B(\H)$\nobreakdash-\hspace{0pt}valued weakly* measurable functions on $X$ with the predual Banach space $L_1(X,\F,\mu_0,\T(\H))$ of $\T(\H)$\nobreakdash-\hspace{0pt}valued Bochner $\mu_{0}$\nobreakdash-\hspace{0pt}integrable functions on $X$. By Theorem 2 in \cite{B&L} with an arbitrary instrument $\M$ one can associate a channel
$\Lambda_{\M}^*:L_\infty(X,\F,\mu_0,\B(\H))\rightarrow\B(\H)$
defined by the relation
$$
\begin{array}{cc}
\displaystyle \Tr \Lambda_{\M}^*(A\otimes f)\rho=\int_X f(x)\Tr\M(dx)[\rho]A,\\\\ A\in\B(\H),\; f\in L_\infty(X,\F,\mu_0),\;\rho\in\S(\H).
\end{array}
$$
The preadjoint channel $\Lambda_{\M}:\T(\H)\rightarrow L_1(X,\F,\mu_0,\T(\H))$ produces the posteriori family as follows
\begin{equation}\label{ps}
\rho_x=\left\{\begin{array}{ll}
  (\Tr\sigma(x))^{-1}\sigma(x), & \Tr\sigma(x)\neq0\\
  \rho_0, & \Tr\sigma(x)=0, \\
\end{array}\right.
\end{equation}
where $\sigma(x)$ is a particular representative of the class $\Lambda_{\M}(\rho)$, while the function $\Tr\sigma(x)$ is a probability density (the Radon-Nikodym derivative) of the measure $\mu_{\rho}$ with respect to the measure $\mu_0$.\smallskip

Consider the channel $\Pi_{\M}^*:L_\infty(X,\F,\mu_0)\rightarrow\B(\H)$ defined by the relation
\begin{equation}\label{Pi}
\Tr\,\rho\,\Pi^*_{\M}(f)=\int_{X}f(x)\Tr\M(dx)[\rho],\;\; f\in L_\infty(X,\F,\mu_0),\;\rho\in\S(\H).
\end{equation}
The preadjoint channel $\Pi_{\M}:\T(\H)\rightarrow L_1(X,\F,\mu_0)$ maps an arbitrary a priori state $\rho$ to the probability density of the posteriori measure $\mu_{\rho}$ with respect to the measure $\mu_0$ and hence it
can be considered as a natural generalization of the channel $\Pi_{\M}$  defined by (\ref{q-c-ch}).

Note that $\,\Pi_{\M}=\Theta\circ\Lambda_{\M}$, where $\Theta$ is the preadjoint channel of the channel
$$
\Theta^*:L_\infty(X,\F,\mu_0)\ni f \mapsto f\otimes I_{\H}\in L_\infty(X,\F,\mu_0,\B(\H)).
$$

Since the channel $\Pi_{\M}$ defined by (\ref{Pi}) has no purely quantum modification, to extend the results of Section 2 we have to generalize the notion of quantum mutual information. \medskip

\begin{definition}\label{m-inf} Let $\A$ be an arbitrary $W^*$-algebra and  $\Phi^*:\A\rightarrow\B(\H)$ be a channel with the preadjoint channel
$\Phi: \T(\H) \rightarrow \A_*$. Let $\rho$ be a state in
$\S(\H)$.  The
\textit{quantum mutual information} of the channel $\Phi$ at the state $\rho$  is defined as follows
\begin{equation*}
I(\rho, \Phi) = H\left(\Phi \otimes \id_{\K}
(\hat{\rho}) \| \Phi \otimes \id_{\K}(\rho\otimes
\varrho)\right),
\end{equation*}
where $\K$ is a Hilbert space isomorphic to $\H$,
$\,\hat{\rho}\,$ is a purification of the state $\rho$ in the space $\H\otimes\K$, $\,\varrho=\Tr_{\H}\hat{\rho}\,$ is a state in $\S(\K)$ isomorphic to $\rho$ and $H(\cdot\|\cdot)$ is the relative entropy for two states in $(\A\otimes\B(\K))_*$.
\end{definition}\medskip

\begin{remark}\label{ls}
It is natural to ask about validity for the above-defined value of the properties of the quantum mutual information of a
purely quantum infinite dimensional channel presented in Propositions 1 and 4 in \cite{H-Sh-3}. Since the proofs of these propositions can not be directly generalized to the case of a channel considered in Definition \ref{m-inf}, the above question is not trivial.

Here we note only that for an arbitrary channel $\Phi:\T(\H)\mapsto\A_*$ the function $\rho\mapsto I(\rho, \Phi)$ is nonnegative and lower semicontinuous and that for an arbitrary channel $\Psi:\A_*\rightarrow\B_*$, where $\B$ is an other $W^*$-algebra, the inequality $I(\rho, \Psi\circ\Phi)\leq I(\rho, \Phi)$ holds for all $\rho$ (the 1-st chain rule). These properties follow from nonnegativity and lower semicontinuity of the relative entropy, Lemma 2 in \cite{H-Sh-3} and Uhlmann's monotonicity theorem \cite{O&P}.
\end{remark}\medskip

Since we will use Definition \ref{m-inf} with $\A=L_{\infty}(X,\F,\mu_0,\B(\H))$, we will deal with the relative entropy for states in
$(\A\otimes\B(\K))_*=L_1(X,\F,\mu_0,\T(\H\otimes\K))$.\smallskip

The relative entropy for two states $\sigma_1$ and $\sigma_2$ in $L_1(X,\F,\mu_0,\T(\H))$ can be expressed as follows
\begin{equation}\label{re}
\begin{array}{cc}
\displaystyle H(\sigma_1\,\|\,\sigma_2)=\int_X \Tr\left(\sigma_1(x)\left(\log\sigma_1(x)-\log\sigma_2(x)\right)\right)\mu_{0}(dx)\\\\\displaystyle=\int_X H_q(\sigma_1(x)\,\|\,\sigma_2(x))\mu_{0}(dx)\\\\\displaystyle=\int_X H_q\left(\frac{\sigma_1(x)}{\Tr\sigma_1(x)}\,\right.\left\|\,\frac{\sigma_2(x)}{\Tr\sigma_2(x)}\right)\mu_{1}(dx)+H_c(\mu_{1}\,\|\,\mu_{2}),
\end{array}
\end{equation}
where $\mu_{1}(dx)=\Tr\sigma_1(x)\mu_{0}(dx)$ (see \cite[formula (4)]{B&L}). In this expression $H_q$ denotes the quantum relative entropy for two positive trace class operators defined by (\ref{qre}), while $H_c$ denotes the classical relative entropy for two probability measures, that is $H_c(\mu_{1}\,\|\,\mu_{2})=\int_X \log\frac{\Tr\sigma_1(x)}{\Tr\sigma_2(x)}\,\mu_{1}(dx)$.
\medskip

For an arbitrary instrument $\M$ equality (\ref{b-eq}) does not hold, but one can prove the following estimation.\medskip
\begin{property}\label{b-eq+}
\emph{Let $\,\M$ be an arbitrary instrument in a Hilbert space $\H$ with the outcome set $\{X,\F\}$ and $\rho$ be a state in $\,\S(\H)$ with finite entropy. Let $\,\hat{\rho}\,$ be a purification of the state $\rho$ in the space $\,\H\otimes\K$. Then
$$
|ER(\rho,\M)-I(\rho, \Pi_{\M})|\leq \int_X H(\hat{\rho}_x)\mu_{\rho}(dx),
$$
where $\,\Pi_{\M}$ is the quantum-classical channel defined by (\ref{Pi}), $\mu_{\rho}(\cdot)=\Tr\M(\cdot)[\rho]$ and $\{\hat{\rho}_x\}$ is the family of posteriori states corresponding to the instrument $\widehat{\M}(\cdot)=\M(\cdot)\otimes\id_{\K}$ and the a priori state $\hat{\rho}$.} \medskip
\end{property}

\textbf{Proof.}  Consider the channels $$\Lambda_{\M}:\T(\H)\rightarrow L_1(X,\F,\mu_0,\T(\H))\quad \textup{and}\quad \Lambda_{\widehat{\M}}:\T(\H\otimes\K)\rightarrow L_1(X,\F,\mu_0,\T(\H\otimes\K))$$ produced by the Barchielli-Lupieri construction described before (since  $\hat{\mu}_{\omega}(\cdot)=\Tr\widehat{\M}(\cdot)[\omega]=\Tr\M(\cdot)[\omega_{\H}]=\mu_{\omega_{\H}}(\cdot)$ for any $\omega\in\S(\H\otimes\K)$, we can use the same measure $\mu_0$ in the both cases). By noting that $L_{\infty}(X,\F,\mu_0,\B(\H\otimes\K))\cong L_{\infty}(X,\F,\mu_0,\B(\H))\otimes\B(\K)$ it is easy to show that $\Lambda_{\widehat{\M}}=\Lambda_{\M}\otimes\id_{\K}$.

Let $\{\hat{\rho}_x\}$ be the family of posteriori states obtained via a given representative of the class $\Lambda_{\widehat{\M}}(\hat{\rho})$ by the rule similar to (\ref{ps}). It is easy to see that $\{\rho_x\doteq\Tr_{\K}\hat{\rho}_x\}$ is the family of posteriori states for the instrument $\M$ corresponding to the a priori state $\rho$. Since the instrument $\widehat{\M}$ is localized in the space $\H$ we have $\int_X \Tr_{\H}\hat{\rho}_x\mu_{\rho}(dx)=\Tr_{\H}\hat{\rho}=\varrho\,\cong\rho$. By using expression (\ref{re}) we obtain
$$
\begin{array}{c}
I(\rho, \Pi_{\M})\doteq H\left(\Pi_{\M}\otimes\id_{\K}(\hat{\rho})\,\|\,\Pi_{\M}(\rho)\otimes\varrho\right)\\\\
=H\left(\Theta\otimes\id_{\K}(\Lambda_{\widehat{\M}}(\hat{\rho}))\,\|\,\Pi_{\M}(\rho)\otimes\varrho\right)
=\int_X H_q(\Tr_{\H}\hat{\rho}_x\,\|\varrho)\mu_{\rho}(dx)\\\\=-\int_X H(\Tr_{\H}\hat{\rho}_x)\mu_{\rho}(dx)+ \int_X \Tr(\Tr_{\H}\hat{\rho}_x(-\log\varrho))\mu_{\rho}(dx)\\\\=-\int_X H(\Tr_{\H}\hat{\rho}_x)\mu_{\rho}(dx)+\Tr\varrho\,(-\log\varrho)
\\\\=[H(\rho)-\int_X H(\rho_x)\mu_{\rho}(dx)]+\int_X (H(\Tr_{\K}\hat{\rho}_x)-H(\Tr_{\H}\hat{\rho}_x))\mu_{\rho}(dx).
\end{array}
$$
By the triangle inequality the absolute value of the last term in this expression is majorized by $\int_X H(\hat{\rho}_x)\mu_{\rho}(dx)$. $\square$ \medskip

Propositions \ref{pure-inst} and \ref{b-eq+} imply the following generalization of equality (\ref{b-eq}).
\smallskip
\begin{corollary}\label{b-eq+c}
\emph{Let $\,\M$ be an efficient instrument in a Hilbert space $\H$.  Then $ER(\rho,\M)=I(\rho, \Pi_{\M})$ for any state  $\rho$ in $\,\S(\H)$ with finite entropy.} \medskip
\end{corollary}

Corollary \ref{b-eq+c} makes it possible to consider the entropy reduction of an efficient quantum measurement not only for a priori states with finite entropy and motivates the following extended version of Definition \ref{QIG}.\medskip

\begin{definition}\label{QIG+}
The entropy reduction of an efficient instrument $\M$ in a Hilbert space $\H$ at an \emph{arbitrary} a priori state $\rho\in\S(\H)$ is
defined as follows
$$
ER(\rho,\M)\doteq I(\rho,\Pi_{\M}),
$$
where $\Pi_{\M}$ is the quantum-classical channel defined by (\ref{Pi}).
\end{definition}\medskip

The following theorem is an extended version of Theorem \ref{QIG-p}.\medskip

\begin{theorem}\label{QIG-p+}
\emph{Let $\,\M$ be an arbitrary efficient instrument in a Hilbert space $\H$. The function $\,\rho\mapsto ER(\rho,\M)$ is nonnegative concave and lower semicontinuous on the set $\,\S(\H)$. It has the following properties:}
\begin{enumerate}[1)]
    \item \emph{$\{ER(\rho,\M)=0\}\Leftrightarrow\{\rho_x\cong\rho\;\,for\; \mu_{\rho}\textup{-}almost\;all\;x\}$,
where $\,\{\rho_x\}$ and $\mu_{\rho}$  are respectively the family of posteriori states and the posteriori probability measure corresponding to the a priori state $\rho$;}

    \item \emph{continuity on any subset of $\,\S(\H)$ on which the von Neumann entropy is continuous:
    $$
\lim_{n\rightarrow+\infty}
H(\rho_n)=H(\rho_0)<+\infty\;\;\Rightarrow\;\;
\lim_{n\rightarrow+\infty}ER(\rho_n,\M)=ER(\rho_0,\M)<+\infty
$$
for any sequence $\{\rho_n\}$ of states converging to a state $\rho_0$;}
\item \emph{monotonicity: for arbitrary efficient instruments $\,\M$ and $\,\N$ in a separable Hilbert space $\H$ the inequality
\begin{equation}\label{monoton+}
ER(\rho, \N\circ\M)\geq ER(\rho, \M)
\end{equation}
holds for any $\,\rho\in\mathfrak{S}(\H);\,$\footnote{The instrument $\,\N\circ\M$ is defined in Lemma \ref{comp}.}}

\item \emph{subadditivity: for arbitrary efficient instruments $\,\M$ and $\,\N$ in separable Hilbert spaces $\H$ and $\K$  the inequality
\begin{equation}\label{subaddit+}
ER(\omega, \M\otimes\N)\leq ER(\omega_{\H},\M)+ER(\omega_{\K},\N)
\end{equation}
holds for any $\,\omega\in\mathfrak{S}(\H\otimes\K)$.\footnote{The instrument $\,\N\otimes\M$ is defined in Lemma \ref{comp}.}}
\end{enumerate}

\end{theorem}\vspace{5pt}

\textbf{Proof.} By Definition \ref{QIG+} lower semicontinuity of the function $\rho\mapsto ER(\rho,\M)$ follows
from the second part of Remark \ref{ls}.

Concavity of the function $\rho\mapsto ER(\rho,\M)$ on the convex subset of states with finite entropy follows from inequality (56) in \cite{B&L}. Concavity of this function on the set $\S(\H)$ can be proved by using lower semicontinuity of this function and Lemma \ref{approx} below (see the proof of Proposition 1 in \cite{H-Sh-3}).\smallskip

Below we will use the notations introduced in the proof of Proposition \ref{b-eq+}. 

1) Note that
$$
ER(\rho,\M)\doteq I(\rho,\Pi_{\M})\doteq H\left(\Theta\otimes \id_{\K}(\Lambda_{\widehat{\M}}(\hat{\rho}))\,\|\,\Theta(\Lambda_{\M}(\rho))\otimes\varrho\right)=0
$$
means
\begin{equation*}
\Theta\otimes \id_{\K}(\Lambda_{\widehat{\M}}(\hat{\rho})) = \Theta(\Lambda_{\M}(\rho))\otimes\varrho
\end{equation*}
by the well known property of the relative entropy. This equality holds if and only if
\begin{equation}\label{is+}
\Tr_{\H}\hat{\rho}_x=\varrho\quad \textup{for}\; \mu_{\rho}\textup{-almost all}\;x.
\end{equation}

"$\Leftarrow$" Let $\varrho_x=\Tr_{\H}\hat{\rho}_x$. Since the instrument $\widehat{\M}$ is irreducible (by Proposition \ref{pure-inst}), we have $\rho_x\cong\varrho_x$ for $\mu_{\rho}$\nobreakdash-\hspace{0pt}almost all $x$ and hence $\varrho_x\cong\varrho=\Tr_{\H}\hat{\rho}$ for $\mu_{\rho}$\nobreakdash-\hspace{0pt}almost all $x$. Since the instrument $\widehat{\M}$ is localized in the space $\H$, we have $\varrho=\int_{X}\varrho_x\mu_{\rho}(dx)$. Thus Lemma \ref{simple} in the Appendix implies (\ref{is+}).\smallskip

"$\Rightarrow$" Since the instrument $\widehat{\M}$ is irreducible (by Proposition \ref{pure-inst}), it follows from (\ref{is+}) that
$\,\rho_x=\Tr_{\K}\hat{\rho}_x\cong\Tr_{\H}\hat{\rho}_x=\varrho\,\cong\rho\,$ for $\mu_{\rho}$-almost all $x$.\smallskip

2) This property follows from identity (\ref{identity}) in Lemma \ref{identity-l}  below, since the both summands in the left side of this identity are lower semicontinuous functions on the set $\S(\H)$ by the second part of Remark \ref{ls}.\smallskip

3) This follows from the 1-st chain rule for the generalized quantum mutual information mentioned in the second part of Remark \ref{ls}. Indeed,
let $\mu_0$ be a measure on $\{X\times Y,\, \F\otimes\E\}$  chosen in accordance with the Barchielli-Lupieri construction for the instrument $\N\circ\M$ and $\nu_0$ be a measure on $\{X, \F\}$ such that $\nu_0(F)=\mu_0(F\times Y)$ for any $F\in\F$. Since $\N(Y)$ is a trace preserving map,
we have $\Tr\N\circ\M(F\times Y)[\rho]=\Tr\N(Y)[\M(F)[\rho]]=\Tr\M(F)[\rho]$ for any $F\in\F$ and $\rho\in\S(\H)$. Hence the measure $\nu_0$ can be used in the Barchielli-Lupieri construction for the instrument $\M$. Let $\Xi$ be a channel from $L_{1}(X\times Y, \F\otimes\E, \mu_0)$ to
$L_{1}(X, \F, \nu_0)$ preadjoint to the channel
$$
L_{\infty}(X, \F, \nu_0)\ni f\mapsto \Xi^{*}(f)=f\otimes \mathbf{1}_y \in L_{\infty}(X\times Y, \F\otimes\E, \mu_0),
$$
where $f\otimes \mathbf{1}_y(x,y)\doteq f(x)\mathbf{1}(y)$. Then $\Pi_{\M}=\Xi\circ\Pi_{\N\circ\M}$. This follows from the relation
$$
\begin{array}{cc}
\Tr\,\rho\,\Pi^*_{\N\circ\M}\circ\Xi^*(f)=\Tr\,\rho\,\Pi^*_{\N\circ\M}(f\otimes \mathbf{1}_y)\doteq
\int_{X\times Y}f(x)\Tr\N(dy)[\M(dx)[\rho]]\\\\=\int_{X}f(x)\Tr\M(dx)[\rho]\doteq\Tr\,\rho\,\Pi^*_{\M}(f),\quad \rho\in \S(\H),\;
f\in L_{\infty}(X, \F, \nu_0),
\end{array}
$$
which can be proved easily by noting that $\N(Y)$ is a trace preserving map. \smallskip

4) Let $\mu_0$ and $\nu_0$ be measures on $\{X,\F\}$ and on $\{Y,\E\}$ chosen in accordance with the Barchielli-Lupieri construction for the instruments $\M$ and $\N$ correspondingly. Then for the instrument $\M\otimes\N$ one can take the measure $\mu_0\otimes\nu_0$. By noting that 
$$
L_{\infty}(X\times Y, \F\otimes\E, \mu_0\otimes\nu_0, \B(\H\otimes\K))=L_{\infty}(X, \F, \mu_0, \B(\H))\otimes L_{\infty}(Y, \E, \nu_0, \B(\K))
$$
it is easy to show that $\Lambda^{*}_{\M\otimes\N}=\Lambda^{*}_{\M}\otimes\Lambda^{*}_{\N}$.

 Let $\omega$ be a state in $\S(\H\otimes\K)$ such that $H(\omega_{\H})$ and $H(\omega_{\K})$ are finite.
Then inequality (\ref{subaddit+}) for the state $\omega$ follows from the inequality
$$
\begin{array}{cc}
H(\omega_{\H})+H(\omega_{\K})-H(\omega)=H(\omega\,\|\,\omega_{\H}\otimes\omega_{\K})\\\\\geq
H(\Lambda_{\M\otimes\N}(\omega)\,\|\,\Lambda_{\M}(\omega_{\H})\otimes\Lambda_{\N}(\omega_{\K}))\\\\\geq
\int_{X\times Y}H_q(\omega_{xy}\,\|\,(\omega_{\H})_x\otimes(\omega_{\K})_y)\mu_{\omega}(dxdy)=
-\int_{X\times Y}H(\omega_{xy})\mu_{\omega}(dxdy)\\\\+\int_{X\times Y}\Tr\,\omega_{xy}\left((-\log(\omega_{\H})_x)\otimes I_{\K}+
 I_{\H}\otimes (-\log(\omega_{\K})_y)\right)\mu_{\omega}(dxdy)\\\\=-\int_{X\times Y}H(\omega_{xy})\mu_{\omega}(dxdy)+\int_{X}H((\omega_{\H})_x)\mu_{\omega_{\H}}(dx)
+\int_{Y}H((\omega_{\K})_y)\mu_{\omega_{\K}}(dy)
\end{array}
$$
obtain by using monotonicity of the relative entropy, expression (\ref{re}) and the equalities
\begin{equation}\label{d-e}
\begin{array}{cc}
\int_{X\times Y}\Tr\,\omega_{xy}((-\log(\omega_{\H})_x)\otimes I_{\K})\mu_{\omega}(dxdy)=\int_{X}H((\omega_{\H})_x)\mu_{\omega_{\H}}(dx),\\\\
\int_{X\times Y}\Tr\,\omega_{xy}(I_{\H}\otimes (-\log(\omega_{\K})_y))\mu_{\omega}(dxdy)=\int_{Y}H((\omega_{\K})_y)\mu_{\omega_{\K}}(dy).
\end{array}
\end{equation}
Prove the first of the above equalities. Let $f$ be a continuous bounded function on $\mathbb{R}$.
Then $f((\omega_{\H})_x)\in L_{\infty}(X, \F, \mu_0, \B(\H))$ and
$$
\begin{array}{cc}
\int_{X\times Y}\Tr\,\omega_{xy}(f((\omega_{\H})_x)\otimes I_{\K})\mu_{\omega}(dxdy)
=\left\langle \Lambda_{\M\otimes\N}(\omega),\, f((\omega_{\H})_x)\otimes (I_{\K}\otimes \mathbf{1}_{y}) \right\rangle\\\\
=\Tr\,\omega \Lambda^*_{\M\otimes\N}(f((\omega_{\H})_x)\otimes (I_{\K}\otimes\mathbf{1}_{y}))=
\Tr\,\omega_{\H} \Lambda^*_{\M}(f((\omega_{\H})_x))=\\\\
=\left\langle \Lambda_{\M}(\omega_{\H}),\, f((\omega_{\H})_x)\right\rangle
=\int_{X}\Tr(\omega_{\H})_x f((\omega_{\H})_x)\mu_{\omega_{\H}}(dx).
\end{array}
$$
Hence the first equality in (\ref{d-e}) can be proved by using approximation of the function $-\log x$ on $[0,1]$ by an increasing sequence
of continuous bounded functions and the monotone convergence theorem.

Let $\omega^0$ be an arbitrary state in $\S(\H\otimes\K)$. Let $\{P_{n}\}$ and $\{Q_{n}\}$
be increasing sequences of  finite rank  spectral projectors of
the states $\omega^0_{\H}$ and $\omega^0_{\K}$ strongly converging to the operators $I_{\H}$ and $I_{\K}$ correspondingly.
Consider the sequence of states
$$\omega^{n}=
\left(\mathrm{Tr}\left((P_{n}\otimes
Q_{n})\cdot\omega^0\right)\right)^{-1}(P_{n}\otimes
Q_{n})\cdot\omega^0\cdot (P_{n}\otimes Q_{n}),
$$
converging to the state $\omega^0$. A direct verification shows that
$$
\lambda_{n}\omega^{n}_{\H}\leq\omega^0_{\H}\quad\textup{and}\quad\lambda_{n}\omega^{n}_{\K}\leq\omega^0_{\K},\quad
\textup{where}\quad \lambda_{n}=\mathrm{Tr}\left((P_{n}\otimes
Q_{n})\cdot\omega^0\right).
$$

Hence concavity and lower semicontinuity of the entropy reduction imply
$$
\lim_{n\rightarrow+\infty} ER(\omega^{n}_{\H},\M)=ER(\omega^0_{\H},\M)\quad \textup{and}\quad \lim_{n\rightarrow+\infty} ER(\omega^{n}_{\K}, \N)=ER(\omega^0_{\K}, \N)
$$
(this can be shown by using the arguments from the proof of \cite[Lemma 6]{Sh-11}).

Since inequality (\ref{subaddit+}) holds with $\omega=\omega^n$ for all $n$, these limit relations
and lower semicontinuity of the entropy reduction show that inequality (\ref{subaddit+}) holds for the state $\omega^0$. $\square$ \medskip

\begin{lemma}\label{approx}
\emph{Let $\,\M$ be an efficient instrument in a Hilbert space $\H$ and $\rho_0$ be
a state in $\,\S(\H)$ with the spectral representation
$\,\rho_{0}=\sum_{i=1}^{+\infty} \lambda_i |e_i\rangle\langle e_i|$. Let
$\,\rho_n = c_n^{-1}\sum_{i=1}^n
\lambda_i |e_i\rangle\langle e_i|$, where $\,c_n =
\sum_{i=1}^n \lambda_i\,$
for each $n$, then}
$$
\lim_{n \rightarrow +\infty} ER(\rho_n,\M)= ER(\rho_0,\M).
$$
\end{lemma}
\textbf{Proof.} Let $\K\cong\H\,$ and $P_n = \sum_{i=1}^n |e_i\rangle\langle e_i|$ be a projector in $\K$, $n=1,2...$  Consider the value
$$
\begin{array}{l}
I_n = H(\Pi_{\M}\otimes \id_{\K}(\hat {\rho}_n)\|\Pi_{\M}(\rho_0)\otimes
\varrho_n) \\\\ = H(c_n^{-1}\Pi_{\M}\otimes\Psi_{n}(\hat {\rho}_0))\|c_n^{-1}\Pi_{\M}(\rho_0)\otimes\Psi_{n}(\varrho_0))),
\end{array}
$$
where
$$
\hat{\rho}_0=\sum_{i,j=1}^{+\infty} \sqrt{\lambda_i \lambda_j}
|e_i\rangle\langle e_j| \otimes |e_i\rangle\langle e_j|,\quad
\hat{\rho}_n=c_n^{-1}I_{\H}\otimes P_n\cdot\hat{\rho}_0\cdot I_{\H}\otimes P_n,
$$
$\varrho_0=\Tr_{\H}\hat{\rho}_0$, $\varrho_n=\Tr_{\H}\hat{\rho}_n$ and $\Psi_n(\cdot)=P_n(\cdot)P_n$ is a map from $\T(\K)$ to itself. We will show that
\begin{equation}\label{l-rel}
\lim_{n \rightarrow +\infty} I_n = H\left(\Pi_{\M}\otimes \id_{\K}(\hat {\rho}_0)\,\|\,
\Pi_{\M}(\rho_0) \otimes
\varrho_0\right) = ER(\rho_0,\M).
\end{equation}
Let $\sigma_1(x)$ and $\sigma_2(x)$ be representatives of the classes $\Pi_{\M}\otimes \id_{\K}(\hat {\rho}_0)$ and
$\Pi_{\M}(\rho_0)\otimes
\varrho_0$ correspondingly. Then $c_n^{-1}P_n\sigma_1(x)P_n$ and $c_n^{-1}P_n\sigma_2(x)P_n$ are respectively representatives of the classes $c_n^{-1}\Pi_{\M}\otimes\Psi_{n}(\hat {\rho}_0)$ and $c_n^{-1}\Pi_{\M}(\rho_0)\otimes\Psi_{n}(\varrho_0)$. Expression (\ref{re}) implies
$$
\begin{array}{cc}
I_n=\int_{X}c_n^{-1}H_q(P_n\sigma_1(x)P_n\,\|\,P_n\sigma_2(x)P_n)\mu_0(dx),\\\\H\left(\Pi_{\M}\otimes \id_{\K}(\hat {\rho}_0)\,\|\,
\Pi_{\M}(\rho_0) \otimes
\varrho_0\right)=\int_{X}H_q(\sigma_1(x)\,\|\,\sigma_2(x))\mu_0(dx)
\end{array}
$$
Hence (\ref{l-rel}) follows from  Lemma 4 in \cite{L-2} and the monotone convergence theorem.

By using expression (\ref{re}) we obtain
$$
0\leq I_n - ER(\rho_n,\M) = H_c(\mu_{\rho_n}\|\,\mu_{\rho_0})=\int_{X}\log\frac{\mu_{\rho_n}(dx)}{\mu_{\rho_0}(dx)}\,\mu_{\rho_n}(dx)\leq-\log c_n
$$
since $c_n\mu_{\rho_n}(F)\leq\mu_{\rho_0}(F)$ for all $F\in\F$.
Hence $\lim_{n}\,(I_n - ER(\rho_n,\M))=0$. This  and (\ref{l-rel}) imply the assertion of the lemma. $\square$\medskip

\begin{lemma}\label{identity-l}
\emph{Let $\,\rho$ be a state in  $\,\S(\H)$ such that $H(\rho)<+\infty$. Then}
\begin{equation}\label{identity}
I(\rho,\Pi_{\M})+I(\rho,\Lambda_{\M})=2H(\rho).
\end{equation}
\end{lemma}

\textbf{Proof.} Since the instrument $\widehat{\M}$ is irreducible (by Proposition \ref{pure-inst}) and is localized in the space $\H$ we have $\,H(\hat{\rho}_x)=0\,$ for $\,\mu_{\rho}$-almost all $\,x\in X$ and $\int_X \Tr_{\H}\hat{\rho}_x\mu_{\rho}(dx)=\Tr_{\H}\hat{\rho}=\varrho\,\cong\rho$. By using expression (\ref{re}) we obtain
$$
\begin{array}{c}
I(\rho, \Lambda_{\M})\doteq H\left(\Lambda_{\widehat{\M}}(\hat{\rho})\,\|\,\Lambda_{\M}(\rho)\otimes\varrho\,\right)=
\int_X H_q(\hat{\rho}_x\,\|\,\rho_x\otimes\varrho\,)\mu_{\rho}(dx)\\\\=-\int_X H(\hat{\rho}_x)\mu_{\rho}(dx)+\int_X \Tr_{\K}\hat{\rho}_x(-\log\rho_{x})\mu_{\rho}(dx)+\int_X \Tr_{\H}\hat{\rho}_x(-\log\varrho\,)\mu_{\rho}(dx)\\\\=\int_X \Tr\rho_x(-\log\rho_{x})\mu_{\rho}(dx)+\Tr\varrho\,(-\log\varrho\,)=
\int_X H(\rho_x)\mu_{\rho}(dx)+H(\rho).
\end{array}
$$
This expression and Corollary \ref{b-eq+c} imply (\ref{identity}). $\square$\medskip

\begin{remark}\label{identity-r}
By proving concavity of the function $\rho\mapsto I(\rho,\Lambda_{\M})$ and by using Lemma 6 in \cite{Sh-11} one can show validity of equality
(\ref{identity}) for any $\rho$ in $\S(\H)$. By comparing this equality with the assertion of Theorem 1 in \cite{H-Sh-3} we see that the channel $\Lambda_{\M}$ plays the role of the complementary channel to the channel $\Pi_{\M}$. Strictly speaking, this holds in the discrete case when the q-c channel $\Pi_{\M}$ can be considered as a purely quantum channel (see Section 2).
\end{remark}\medskip

The following proposition is a generalization of Proposition \ref{cont-cond}.\smallskip

\begin{property}\label{cont-cond+}
\emph{Let $\,\{\M_n\}$ be a sequence of efficient quantum instruments with the same outcome space $\{X,\F\}$ converging to the instrument $\,\M_0$ in the following sense
\begin{equation}\label{ins-conv-1}
\|\cdot\|_1\,\textup{-}\!\lim_{n\rightarrow+\infty}\M_n(F)[\rho]=\M_0(F)[\rho] \quad \forall F\in\F,\; \forall\rho\in\S(\H).
\end{equation}
Then for an arbitrary sequence $\{\rho_n\}$ of states in $\,\S(\H)$ converging to a state $\rho_0$ the following relation holds
$$
\liminf_{n\rightarrow+\infty}ER(\rho_n,\M_n)\geq ER(\rho_0,\M_0).
$$
If, in addition, $\,\lim_{n\rightarrow+\infty}H(\rho_n)=H(\rho_0)<+\infty\,$ then $$\lim_{n\rightarrow+\infty}ER(\rho_n,\M_n)= ER(\rho_0,\M_0).$$}
\end{property}
\textbf{Proof.} By Lemma 2 in \cite{H-Sh-3} there exists a sequence $\{\hat{\rho}_n\}$ of purifications of the states $\{\rho_n\}$ converging to a   purification $\hat{\rho}_0$ of the state $\rho_0$. By using Lemma \ref{cont-cond+l} below it is easy to show that
$p.w.\textup{-}\lim_{n\rightarrow+\infty}\Lambda_{\M_n}(\rho_n)=\Lambda_{\M_0}(\rho_0)$ and
$p.w.\textup{-}\lim_{n\rightarrow+\infty}\Lambda_{\widehat{\M}_n}(\hat{\rho}_n)=\Lambda_{\widehat{\M}_0}(\hat{\rho}_0)$.

The first assertion follows from lower semicontinuity of the relative entropy with respect to pointwise convergence of states \cite[Corollary 5.12]{O&P}.

The second assertion follows from  identity (\ref{identity}) in Lemma \ref{identity-l}, since by the above arguments we have
$$
\liminf_{n\rightarrow+\infty}I(\rho_n,\Pi_{\M_n})\geq I(\rho_0,\Pi_{\M_0})\quad\textup{and}\quad \liminf_{n\rightarrow+\infty}I(\rho_n,\Lambda_{\M_n})\geq I(\rho_0,\Lambda_{\M_0}).\;\square
$$

Proposition \ref{cont-cond+} implies the following "continuous" version of the third assertion of Corollary 2 in \cite{H-Sh-3}.
Let
$\,\langle H\hspace{1pt}\rangle_{\M,\rho}\doteq\int_X H(\rho_x)\mu_{\rho}(dx)\,$ be the mean entropy of posteriori states corresponding
to a quantum instrument $\M$ and an a priory state $\rho$.\smallskip

\begin{corollary}\label{cont-cond+c}
\emph{Let $\,\{\M_n\}$ be a sequence of efficient quantum instruments with the same outcome space $\{X,\F\}$ converging to the instrument $\,\M_0$ in the sense of (\ref{ins-conv-1}) and $\{\rho_n\}$ be a sequence in $\,\S(\H)$ converging to a state $\rho_0$ such that $\;\lim_{n\rightarrow+\infty}H(\rho_n)=H(\rho_0)<+\infty$. Then}
$$
\lim_{n\rightarrow+\infty}\langle H\hspace{1pt}\rangle_{\M_n,\rho_n}=\langle H\hspace{1pt}\rangle_{\M_0,\rho_0}.
$$
\end{corollary}

It is easy to see that there exists a positive complete measure $\mu_{0}$ on $(X,\F)$ which can be used in the Barchielli-Lupieri construction for each instrument from the sequence $\,\{\M_n\}$. \smallskip

\begin{lemma}\label{cont-cond+l}
\emph{Convergence of the sequence $\{\M_n\}$ to the instrument $\M_0$ defined by (\ref{ins-conv-1}) means that
\begin{equation}\label{ins-conv-2}
   p.w.\,\textup{-}\!\lim_{n\rightarrow+\infty}\Lambda_{\M_n}(\rho)=\Lambda_{\M_0}(\rho)\quad\forall\rho\in\S(\H),
\end{equation}
that is $\,\lim_{n\rightarrow+\infty}\langle\Lambda_{\M_n}(\rho),\widehat{A}\rangle=\langle\Lambda_{\M_0}(\rho),\widehat{A}\rangle$ for any
$\widehat{A}\in L_{\infty}(X, \F, \mu_0, \B(\H))$.}
\end{lemma}\smallskip

\textbf{Proof.} Note first that $\,\lim_{n\rightarrow+\infty}\langle\Lambda_{\M_n}(\rho),\chi_F\otimes A\rangle=\langle\Lambda_{\M_0}(\rho),\chi_F\otimes A\rangle$, where $\chi_F$ is the indicator function of the set $F\in\F$ and $A\in\B(\H)$, means that $\lim_{n\rightarrow+\infty}\Tr\M_n(F)[\rho]A=\Tr\M_0(F)[\rho]A$. Thus (\ref{ins-conv-2}) implies (\ref{ins-conv-1}).

To prove the converse implication assume that $\H_0$ is a finite-dimensional subspace of $\H$. Since $L_{\infty}(X, \F, \mu_0, \B(\H_0))$ coincides with the $C^*$-tensor product of
$L_{\infty}(X, \F, \mu_0)$ and $\B(\H_0)$, arbitrary $\widehat{A}_0\in L_{\infty}(X, \F, \mu_0, \B(\H_0))$ can be approximated in the norm topology by a sequence $\{\widehat{A}_m\}$ belonging to the linear span of the set $\{\chi_F\otimes A\,|\,F\in\F,A\in\B(\H_0)\}$. As mentioned before
(\ref{ins-conv-1}) implies $\lim_{n\rightarrow+\infty}\langle\Lambda_{\M_n}(\rho),\widehat{A}_m\rangle=\langle\Lambda_{\M_0}(\rho),\widehat{A}_m\rangle$  for each $m$. By using the standard argumentation we conclude that $\,\lim_{n\rightarrow+\infty}\langle\Lambda_{\M_n}(\rho),\widehat{A}_0\rangle=\langle\Lambda_{\M_0}(\rho),\widehat{A}_0\rangle$.

Note that (\ref{ins-conv-1}) implies that the set $\{\M_n(X)[\rho]\}_{n\geq0}$ of states in $\S(\H)$ is compact. By the compactness criterion for subsets of $\S(\H)$ (see Lemma 10 in \cite{Sh-11}) for arbitrary $\varepsilon>0$ there exists a finite dimensional projector $P_\varepsilon$ such that $\Tr\M_n(X)[\rho]P^{\perp}_{\varepsilon}<\varepsilon$ for all $n=0,1,2,...$, where $P^{\perp}_{\varepsilon}=I_{\H}-P_{\varepsilon}$. This means that
\begin{equation}\label{comp-imp}
\int_X \Tr P^{\perp}_{\varepsilon}\sigma_n(x)\mu_0(dx)<\varepsilon\quad \textup{for}\;\textup{all}\;\,n=0,1,2,...,
\end{equation}
where $\sigma_n(x)$ is a representative of the class $\Lambda_{\M_n}(\rho)$.

Let $\varepsilon>0$ be arbitrary and $A(x)$ be a representative of a class $\widehat{A}\in L_{\infty}(X, \F, \mu_0, \B(\H))$. Then
$$
\begin{array}{cc}
 \langle\Lambda_{\M_n}(\rho),\widehat{A}\rangle=\int_X \Tr A(x)\sigma_n(x)\mu_0(dx)\\\\=
\int_X \Tr P_\varepsilon A(x)P_\varepsilon\sigma_n(x)\mu_0(dx)+\int_X \Tr P^{\perp}_{\varepsilon} A(x)P_\varepsilon\sigma_n(x)\mu_0(dx)\\\\+\int_X \Tr P_\varepsilon A(x)P^{\perp}_{\varepsilon}\sigma_n(x)\mu_0(dx)+\int_X \Tr P^{\perp}_{\varepsilon} A(x)P^{\perp}_{\varepsilon}\sigma_n(x)\mu_0(dx).
\end{array}
$$
By means of (\ref{comp-imp}) it is easy to show that the last three terms in this expression are less than $\varepsilon\|\widehat{A}\|$ for all $n=0,1,2,...$ As proved before (\ref{ins-conv-1}) implies that the first term tends to $\int_X \Tr P_\varepsilon A(x)P_\varepsilon\sigma_0(x)\mu_0(dx)$, since
$P_\varepsilon A(x)P_\varepsilon\in\B(P_\varepsilon(\H))$. Thus we conclude that (\ref{ins-conv-2}) holds. $\square$

\section{Appendix}

\subsection {A characterization of a quantum instrument which is irreducible but not efficient}

In Theorem 1 in \cite{H-RND} the representation of a quantum instrument analogous to the Kraus representation of a completely positive map is obtained. By this theorem for an arbitrary instrument $\M$ in a Hilbert space $\H$ with the outcome set $\{X,\,\F\}$ there exists a positive $\sigma$-finite measure $\mu$ on $\,\{X,\,\F\}$, a dense domain $\D\subset\H$ and a countable family of functions $x\mapsto V_k(x)$ defined for $\mu$-almost all $x$, such that $V_k(x)$ is a linear operator from $\D$ to $\H$, satisfying
$$
\langle\varphi|\M^*(F)[A]\varphi\rangle=\int_F\sum_k\langle V_k(x)\varphi|AV_k(x)\varphi\rangle\mu(dx)
$$
for any $\,\varphi\in\D$, $\,F\in\F$, $\,A\in\B(\H)$. We may assume that for each $x$ the all nonzero operators from the family $V_k(x)$ are not    proportional to each other, since if $\,V_{k'}(x_0)=\lambda V_k(x_0)$ for some $x_0$ and $\lambda\in\mathbb{C}$ then we may replace $V_k(x_0)$  by $\sqrt{1+|\lambda|^2}V_k(x_0)$ and consider that $V_{k'}(x_0)=0$.

The instrument $\M$ is efficient if and only if  $V_k(x)=0$ for $k>1$ and $\mu$-almost all $x$. This follows
from the proof of Theorem 1 in \cite{H-RND} and the proof of the implication $\mathrm{(iv)\Rightarrow(i)}$ in Proposition \ref{pure-inst}.

So, if the instrument $\M$ is not efficient then there exists a subset $F_s$ of $X$ such that $\M(F_s)\neq0$ and
$V_2(x)\neq0$ for all $x\in F_s$.\footnote{We assume that if $V_k(x)=0$ then $V_{k'}(x)=0$ for all $k'>k$.} \medskip

\begin{property}\label{i-np-ch}
\emph{If the instrument $\;\M$ is irreducible but not efficient then for $\mu$-almost all $x$ in $F_s$ the all nonzero operators $V_k(x)$ has the same one dimensional range (depending on $x$), that is $V_k(x)|\varphi\rangle=\omega_k(x)[\varphi]|\psi_x\rangle$ for all $\varphi\in\D$, where $\omega_k(x)$ is a linear functional defined on $\D$ (not necessary bounded) and $\psi_x$ is a unit vector in $\H$ (not depending on $\varphi$). This means that
\begin{equation}\label{i-np-ch+}
\langle\varphi|\M^*(F)[A]\varphi\rangle=\int_F\langle \psi_x|A\psi_x\rangle\sum_k|\omega_k(x)[\varphi]|^2\mu(dx)
\end{equation}
for any $\,\varphi\in\D$, $\,A\in\B(\H)$ and $\,F\in\F$ such that $\,F\subseteq F_s$.}

\emph{It follows that for an arbitrary a priori state $\rho$ the posteriori state is $\rho_x=|\psi_x\rangle\langle\psi_x|\,$ for $\,\mu_{\rho}$-almost all $\,x\in F_s$  (that is, $\rho_x$ does not depend on $\,\rho$).}
\end{property}\medskip

This assertion can be obtained by using the arguments from the proof of the implication $\mathrm{(ii)\Rightarrow(iv)}$ in Proposition \ref{pure-inst} with $\dim\K=1$. The only difference appears at the point where Lemma \ref{la} is used, since in this case we can not exclude the possibility of $\mathrm{rank}V_k(x)=1$. It is this possibility that prevents to prove that any irreducible instrument is efficient. \medskip

\begin{corollary}\label{i-np-ch-c}
\emph{If $\,\M$ is an instrument not taking values within the set of nonzero entang-lement-breaking quantum operations\footnote{A quantum operation $\Phi$ is called entanglement-breaking if for an arbitrary state $\omega$ in $\S(\H\otimes\K)$, where $\K$ is a separable Hilbert space, the operator $\Phi\otimes\id_{\K}(\omega)$ belongs to the convex closure of the product-operators $A\otimes B$, $A\in\T_{+}(\H)$, $B\in\T_{+}(\K)$.}
then $\,\M$ is efficient if and only if $\,\M$ is irreducible.}
\end{corollary}\medskip

\textbf{Proof.} Let $\M$ be a instrument in a Hilbert space $\H$ with the outcome set $\{X,\,\F\}$, which is irreducible but not efficient. By Proposition \ref{i-np-ch} there exists a positive $\sigma$-finite measure $\mu$ on $\,\{X,\,\F\}$, a subset $F$ of $X$ such that $\M(F)\neq0$, a dense domain $\D\subseteq\H$, a family $\{x\mapsto\omega_k(x)\}_k$ of functions on $F$, such that $\omega_k(x)$ is a linear functional defined on $\D$ for each $x\in F$, and a function $x\mapsto\psi_x$ on $F$ such that $\psi_x$ is a unit vector in $\H$ for each $x\in F$, for which relation (\ref{i-np-ch+}) holds. By using the polarization identity one can show that
\begin{equation}\label{i-np-ch++}
\M(F)[|\varphi_1\rangle\langle\varphi_2|]=\int_F|\psi_x\rangle\langle\psi_x|\sum_k\omega_k(x)[\varphi_1]\,\overline{\omega_k(x)[\varphi_2]}\mu(dx),
\end{equation}
where $\int$ denotes the Bochner integral.

Let $\omega=\sum_{i,j=1}^{m}|\varphi_i\rangle\langle\varphi_j|\otimes|\phi_i\rangle\langle\phi_j|$ be a pure state in $\S(\H\otimes\K)$, where
$\{\varphi_i\}_{i=1}^{m}\subset\D$ and $\{\phi_i\}_{i=1}^{m}\subset\K$. Then (\ref{i-np-ch++}) implies
$$
\begin{array}{cc}
\displaystyle\M(F)\otimes \id_{\K}(\omega)=\int_F|\psi_x\rangle\langle\psi_x|\otimes\sum_k\sum_{i,j=1}^{m}\omega_k(x)[\varphi_i]\,\overline{\omega_k(x)[\varphi_j]}
\,|\phi_i\rangle\langle\phi_j|\,\mu(dx)\\\\\displaystyle=\int_F|\psi_x\rangle\langle\psi_x|\otimes\sum_k|\eta^k_x\rangle\langle\eta^k_x|\,\mu(dx),
\end{array}
$$
where $|\eta^k_x\rangle=\sum_{i=1}^{m}\omega_k(x)[\varphi_i]|\phi_i\rangle$ is a vector in $\K$. It follows that the operator $\M(F)\otimes \id_{\K}(\omega)$ is separable. Since an arbitrary pure state in $\S(\H\otimes\K)$ can be approximated by a sequence of above-considered states, the operator $\M(F)\otimes \id_{\K}(\omega)$ is separable for any state $\omega\in\S(\H\otimes\K)$. Thus the operation $\M(F)$ is  entanglement-breaking. $\square$ \medskip

\begin{example}\label{i-np-ch-e}
Let $P$ be a spectral projector valued measure on a measurable space $\{X,\F\}$ and $|\psi_0\rangle$ be a fixed unit vector in $\H$. Then the instrument $\,\M(F)[\rho]=[\Tr P(F)\rho]|\psi_0\rangle\langle\psi_0|\,$ is obviously irreducible, but it is efficient if and only if the spectral measure $P$ has (uniform) multiplicity one, since it is easy to see that the spectral measure from representation (\ref{SR}) coincides with $P$.
Note that the channel $\,\M(X):\rho\mapsto|\psi_0\rangle\langle\psi_0|\,$ is entanglement-breaking.
\end{example}

\subsection{Two auxiliary lemmas}

\begin{lemma}\label{la}
\emph{Let $\,\L_1=\mathrm{lin}(\{\varphi_i\}_{i\in\mathbb{N}})$ be a linear space and  $\L^0_1$ be a countable subset of $\,\L_1$ consisting of finite linear combinations of the vectors $\,\varphi_1, \varphi_2, ...$ with rational coefficients. Let $\,\{A_k\}$ be a finite or countable family of nonzero linear operators from
$\,\L_1$ to a linear space $\,\L_2$ such that the set $\,\{A_k(\varphi)\}\subset\L_2$ consists of collinear vectors for any $\varphi\in\L^0_1$. If at least one operator
in the family $\,\{A_k\}$ has rank $>1$ then $A_k=\lambda_k A_1$ for all $\,k$, where $\,\{\lambda_k\}$ is a set of nonzero scalars.}
\end{lemma}\smallskip

\textbf{Proof.} Suppose $\mathrm{rank}A_1>1$ and $k$ is arbitrary. By the condition $A_k(\varphi)=\lambda_k^\varphi A_1(\varphi)$ for all $\varphi\in\L^0_1\setminus\ker A_1$. We will show that $\lambda_k^\varphi$ does not depend on $\varphi$.

Since $\mathrm{rank}A_1>1$, without loss of generality we may assume that the vectors $A_1(\varphi_1)$ and $A_1(\varphi_2)$ are not collinear. Let $\psi=c_1\varphi_1+c_2\varphi_2$, where $c_1$ and $c_2$ are nonzero rational coefficients. By linearity we have
$$
A_k(\psi)=\lambda_k^\psi A_1(\psi)=\lambda_k^\psi (c_1 A_1(\varphi_1)+c_2 A_1(\varphi_2))
$$
and
$$
A_k(\psi)=c_1 A_k(\varphi_1)+c_2 A_k(\varphi_2)= c_1 \lambda_k^{\varphi_1} A_1(\varphi_1)+c_2 \lambda_k^{\varphi_2} A_1(\varphi_2).
$$
Hence $\lambda_k^{\varphi_1}=\lambda_k^{\varphi_2}$.  Let $\varphi$ be an arbitrary vector in $\L^0_1\setminus\ker A_1$.
Then the vector $A_1(\varphi)$ is not collinear with $A_1(\varphi_i)$, where either $i=1$ or $i=2$. By repeating the above arguments for the pair $(\varphi,\varphi_i)$ instead of $(\varphi_1,\varphi_2)$ we obtain $\lambda_k^{\varphi}=\lambda_k^{\varphi_i}$.
Thus $\lambda_k^\varphi=\lambda_k$ for all $\varphi\in\L^0_1\setminus\ker A_1$.

If $\lambda_k=0$ then $A_k(\varphi)=0$ for all $\varphi\in\L^0_1\setminus\ker A_1$. This implies  $A_k=0$ contradicting to the assumption. Indeed, if $\varphi_i\in\ker A_1$ for some $i$ then $\varphi_i=(\varphi_i+\varphi_1)-\varphi_1$ and hence $A_k(\varphi_i)=A_k(\varphi_i+\varphi_1)-A_k(\varphi_1)=0$.

Thus we have
\begin{equation}\label{one}
 A_k(\varphi)=\lambda_k A_1(\varphi)\;\, \textup{for}\; \textup{all}\;\, \varphi\in\L^0_1\setminus\ker A_1,\;\, \textup{where}\; \lambda_k\neq0.
\end{equation}
Hence the vectors $A_k(\varphi_1)$ and $A_k(\varphi_2)$ are not collinear.
By repeating the above arguments with $A_k$ instead of $A_1$ and $A_1$ instead of $A_k$  we obtain
\begin{equation}\label{two}
 A_1(\varphi)=\lambda'_k A_k(\varphi)\;\, \textup{for}\; \textup{all}\;\, \varphi\in\L^0_1\setminus\ker A_k,\;\, \textup{where}\; \lambda'_k\neq0.
\end{equation}
It follows from (\ref{one}) and (\ref{two}) that $A_k(\varphi_i)=\lambda_k A_1(\varphi_i)$ for all $i\in\mathbb{N}$ and hence
$A_k=\lambda_k A_1$. $\square$ \medskip

\begin{lemma}\label{simple} \textit{Let $\,\{\pi_\alpha, \rho_\alpha\}$ be a countable or continuous ensemble of states in  $\S(\H)$ such that $\rho_\alpha\cong\bar{\rho}$ for all $\,\alpha$, where $\bar{\rho}$ is the average state of this ensemble. Then $\,\rho_\alpha=\bar{\rho}\,$ for all $\,\alpha$.}
\end{lemma}\smallskip

The assertion of this lemma follows from existence of a finite strictly convex function on the set $\S(\H)$ depending only on the spectrum of a state. As the simplest example one can consider the function $f(\rho)=\Tr\rho^2$.
\medskip

I am grateful to A.S.Holevo and the participants of his seminar "Quantum probability, statistic, information" for useful discussion. I am
also grateful to the organizers of the program "Quantum information theory" in the Institut Mittag-Leffler (Djursholm, Sweden), where the work was completed, and to the participants of this program, especially, to E.Effros, M.Wolf and C.Palazuelos for the help in solving the particular problems.

This work is partially supported by the program
"Mathematical control theory" of Russian Academy of Sciences, by the
federal target program "Scientific and pedagogical staff of
innovative Russia" (program 1.2.1, contract P 938) and by
RFBR grant 09-01-00424-a.

\end{document}